\documentclass[twocolumn,preprintnumbers,amsmath,amssymb]{revtex4-2}
\usepackage{graphicx}  
\usepackage{sidecap}
\begin{document}
%
\title{Terahertz  plasmonic resonances in coplanar graphene nanoribbon structures
}
\author{V.~Ryzhii$^{1,*}$, C.~Tang$^{1,2}$, T.~Otsuji$^1$, M.~Ryzhii$^{3}$,
 and M.~S.~Shur$^4$}
\address{
$^1$Research Institute of Electrical Communication,~Tohoku University,~Sendai~ 980-8577, 
Japan\\
$^2$Frontier Research Institute for Interdisciplinary Studies, Tohoku University, Sendai 980-8577, Japan\\
$^3$Department of Computer Science and Engineering, University of Aizu, Aizu-Wakamatsu 965-8580, Japan\\
$^4$Department of Electrical,~Computer,~and~Systems~Engineering, Rensselaer Polytechnic Institute,~Troy,~New York~12180,~USA\\
* {\bf Author to whom correspondence should be addressed:} v-ryzhii@gmail.com
}
\makeatother


\date{\today}

\begin{abstract}
\normalsize We analyze plasmonic oscillations in the coplanar graphene nanoribbon (GNR) structures
induced by the applied terahertz (THz) signals and calculate the GNR impedance. The
plasmonic oscillations in the GNR structures are associated with the electron and hole inductances
and the lateral inter-GNR capacitance. A relatively low inter-GNR capacitance
enables the resonant excitation of the THz plasmonic oscillations in the GNR structures with
long GNRs. The GNR structures under consideration can be used in different THz devices as
the resonant structures incorporated in THz detectors, THz sources using resonant-tunneling
diodes, photomixers, and surface acoustic wave sensors.
\end{abstract}
\maketitle

\section{Introduction}

Due to the remarkable properties of graphene layers (GLs) (high intrinsic mobility enabling an elevated 
electrical conductivity, high  thermal conductivity, transparency to light, and mechanical strength), GLs are very attractive
 for  different device applications. In particular, graphene micro- and nanoribbons 
can be used in the  detectors and sources of electromagnetic radiation, 
surface 
acoustic wave sensors, power and flexible devices
~\cite{1,2,3,4,5,6,7,8,9,10,11,12,13,14}. 
The possibility of the terahertz (THz) plasmonic wave  
excitation in  the graphene structures~\cite{2,3,4,7,8,9,10,15,16,17,18,19,20,21,22,23} markedly enhances the functionality of the graphene-based devices.
The structures with GLs partitioned into arrays of graphene nanoribbons (GNRs) have also been already studied (for example,~\cite{24,25,26,27,28,29}). In this work, we focused mainly on the plasmonic waves
propagating perpendicular to the GNRs aiming the achievement of the plasmonic resonances at relatively high frequencies, particularly, in the frequencies corresponding to the  mid-infrared range.

In the present paper,
we explore the structures comprising two co-planar GNRs on a dielectric substrate
focusing on their terahertz applications. In such structures the plasmonic wave number is determined by
the GNR length, which is much longer than their widths. As a result, the perpendicular plasmonic modes
are not excited in the frequency range under consideration (the THz range).
 
 It is assumed that the bias dc voltage $V_G$ and the THz signal
voltage $\delta V_{\omega}$  with the frequency $\omega$ are applied between the GNR edges.
The bias voltage results in the formation of the two-dimensional electron and hole systems,
2DES and 2DHS. The signal voltage can be produced by the impinging electromagnetic radiation  and received by an antenna integrated with the GNR structure. The THz signals excite the plasmonic standing waves (plasmonic oscillations) 
in the GNRs.  Examining the plasmonic properties of the co-planar GNRs,
we show that with a relatively low geometrical inter-GNR capacitance,
the plasmonic resonances can be in the THz range even for fairly long GNRs.
This property of the  coplanar GNRs, as well as their lateral periodic arrays, can find  applications in different  devices
using the THz plasmonic resonant structures.

\section{Equations of the model}

\begin{figure}[t]
\centering
\includegraphics[width=9.0cm]{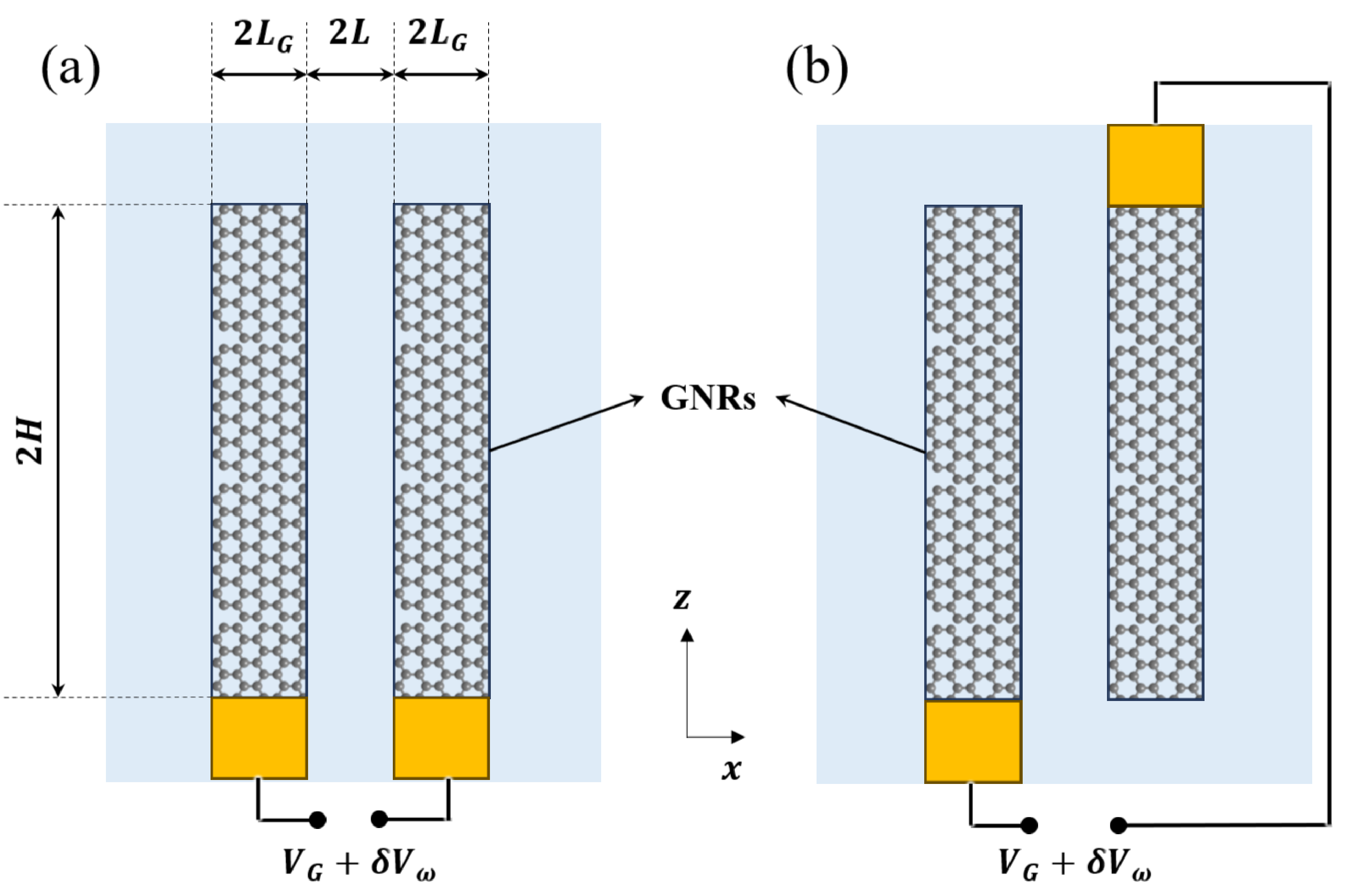}
\caption{Schematic top view of  the co-planar GNR structures with different types of  voltage bias.}
\label{Fig1}
\end{figure}

Considering the co-planar GNR structures, we assume 
two types of  signal voltage bias, P-GNR and S-GNR,   as shown in Fig.~1(a) and Fig.~1(b).  The lengths and widths  of the GNRs are $2H$ and $2L_G$, respectively, whereas the spacing between the GNRs is equal to $2L$ being much smaller than other
geometrical parameters ($L_G, L \ll H$ ).

The  GNR potentials are $\varphi(z,t)^{\pm} =\delta\varphi_{\omega}^{\pm}(z)e^{-i\omega t}$,
where the signs "+" and "-" correspond to the GNRs to which the voltages of different polarities are applied.
The displacement  current  density between the  GNRs and
the continuity equations for the electrons and holes governing their transport along the GNRs 
can be presented as (see Appendix)

\begin{eqnarray}\label{eq1}
\delta j_{\omega} =  -i\omega\,C(\delta\varphi_{\omega}^{+}-\delta\varphi_{\omega}^{-}),
\end{eqnarray} 
\begin{eqnarray}\label{eq2}
 2L_G\sigma_{G,\omega}\frac{d^2 \delta\varphi_{\omega}^{\pm}}{d z^2} = 
\mp i\omega\,C(\delta\varphi_{\omega}^{+}-\delta\varphi_{\omega}^{-}).
\end{eqnarray} 
Here
\begin{eqnarray}\label{eq3}
 \sigma_{G,\omega} = \sigma_G\frac{i\nu}{(\omega+i\nu)}
\end{eqnarray} 
and

\begin{eqnarray}\label{eq4}
 C=\frac{\kappa}{2\pi^2}c
 \end{eqnarray} 
are the GNR ac conductivity and the geometrical inter-GNR capacitance, respectively,
where $\sigma_G = e^2\mu/\pi\hbar^2\nu$ is the dc Drude conductivity of a GNR with the carrier Fermi energy $\mu$ and the carrier collision frequency $\nu$ (scattering by impurities and acoustic phonons), $\kappa = (\kappa_S+1)/2$ and $\kappa_S$ are the effective dielectric constant and the substrate dielectric constant (the dielectric constant of the media above the structure is assumed to be unity), 
the factor 
$$
c  = a\tan^{-1}\biggl(\frac{1}{\sqrt{a^2-1}}\biggr) +\ln(a+\sqrt{a^2-1})
$$
reflects the GNR
 blade-like shape~\cite{30} (see also Refs.~\cite {31,32,33,34,35}) with  parameter $a = L_G/L $.

The electron and hole Fermi energy in the pertinent GNR is determined by the bias voltage $V_G$: $\mu \simeq \hbar\,v_W\sqrt{\pi\Sigma_G} \simeq \hbar\,v_W\sqrt{\kappa\,c\,V_G/2\pi\,eL_G}$, where $v_W \simeq 10^8$~cm/s is the characteristic electron and hole velocity in graphene, $\Sigma_G$ the steady-state electron and hole density, and $\hbar$ is the Planck constant.

Using Eqs.~(1) - (3), we arrive at  the following equations governing the potentials $\delta \varphi_{\omega}^{\pm}$:

\begin{eqnarray}\label{eq5}
\frac{\partial^2 \delta \varphi_{\omega}^{+}}{\partial z^2} \pm
\frac{(\omega + i\nu)\omega}{s^2}
(\delta \varphi_{\omega}^{+} - \delta \varphi_{\omega}^{-}) = 0,
\end{eqnarray}

\begin{eqnarray}\label{eq6}
\frac{\partial^2 \delta \varphi_{\omega}^{-}}{\partial z^2} -
\frac{(\omega + i\nu)\omega}{s^2}
(\delta \varphi_{\omega}^{+} - \delta \varphi_{\omega}^{-}) = 0,
\end{eqnarray} 
with $s = \sqrt{2e^2\mu\,L_G/\pi\,C\hbar^2}$ being the characteristic velocity of the plasmonic wave along the GNRs.
Depending on the signal voltage application, the boundary conditions for Eqs.~(5) and (6) are:

\begin{eqnarray}\label{eq7}
\delta\varphi_{\omega}^{\pm}|_{z= -H} = \pm\delta V_{\omega}/2, \quad \partial\delta\varphi_{\omega}^{\pm}/\partial z|_{z= H} = 0.
\end{eqnarray}
for the situation corresponding to the device corresponding to structure Fig~1(a)  P-GNR, and

\begin{eqnarray}\label{eq8}
\delta\varphi_{\omega}^{\pm}|_{z= \mp H} = \pm\delta V_{\omega}/2,\qquad \partial\delta\varphi_{\omega}^{\pm}/\partial z|_{z= \pm H} = 0.
\end{eqnarray}
for the device shown in Fig.~1(b) - interdigital GNR connection  S-GNR.

\section{Plasmonic oscillations}

Equations~(5) and (6) with the boundary conditions given by Eqs.(7) and (8) for the local signal voltage swing between the GNRs 
 $\delta \varphi_{\omega} = (\delta \varphi_{\omega}^{+} - \delta \varphi_{\omega}^{-})$ 
 result in

 \begin{eqnarray}\label{eq9}
 \delta\varphi_{\omega}^P = 
 \delta V_{\omega}\frac{\cos(\gamma_{\omega}z/H)}{\cos\gamma_{\omega}}
 \end{eqnarray}
and

 \begin{eqnarray}\label{eq10}
 \delta\varphi_{\omega}^S = 
 \delta V_{\omega}\frac{\cos(\gamma_{\omega}z/H)}{[\cos\gamma_{\omega} -\gamma_{\omega}\sin\gamma_{\omega}]}
 \end{eqnarray}
for the P- and S-GNR structures, respectively.
 
Here 
\begin{eqnarray}\label{eq11}
 \gamma_{\omega} =\frac{H\sqrt{(\omega+\nu)\omega}}{S} = \frac{ \pi\sqrt{(\omega+i\nu)\omega}}{2\Omega},
 \end{eqnarray} 
where

\begin{eqnarray}\label{eq12}
\Omega = \frac{e}{H\hbar}\sqrt{\frac{  \pi\mu\,L_G}{4C}}\simeq 
\frac{e}{H\hbar}\sqrt{\frac{\pi^3\mu\,L_G}{2\kappa\,c}}
\end{eqnarray}
is the  characteristic frequency of the plasmonic oscillation  fundamental mode  in
the  GNRs. Considering the dependence of $\mu$ on $V_G$,
we obtain

\begin{eqnarray}\label{eq13}
\frac{\Omega}{2\pi} = \frac{1}{4H}\biggl(\frac{2\pi\,e^3 v_W^2}{\hbar^2
}\frac{L_GV_G}{\kappa\,c}\biggr)^{1/4}\propto
\frac{1}{H}\biggl(\frac{L_GV_G}{\kappa\,c}\biggr)^{1/4}. 
\end{eqnarray}
In particular,  for $\kappa_S = 4$ ($\kappa = 2.5$), $\mu = 140$~meV, $L_G = 0.5~\mu$m, $L=0.25~\mu$m ($c\simeq 2.5$), $H = (2.0 -4.0)~\mu$m, Eq.~(13) yields $\Omega/2\pi \simeq (1.0 - 2.0)$~THz. 

Equations~(9) and (10) describe the spatial distributions   of the self-consistent potential 
and carrier densities in the GNRs (along them), i.e., the plasmonic oscillations (the standing plasmonic waves with the frequency $\omega$ and the normalized wave number $\gamma_{\omega}$).
As follows from Eqs.~(9) and (10), the relative amplitudes of the ac potential $|\delta \varphi_{\omega}^P|/\delta V_{\omega}$ and $|\delta \varphi_{\omega}^S|/\delta V_{\omega}$ can be markedly larger than unity when the denominators in right-hand sides of these equations
$|\cos \gamma_{\omega}| \ll 1$ and  $|\cos \gamma_{\omega} - \gamma_{\omega}\sin \gamma_{\omega}| \ll 1$
at certain signal frequencies (the plasmonic resonant frequencies).

Deriving Eqs.~(5),  (6), (9), and (10), we disregarded the ac potential nonuniformity across the GNRs (in the $x$-direction), i.e., neglected the plasmonic oscillation longitudinal modes with the wave vectors $k_x \propto 1/L_G$ directed perpendicular to the GNR. Such modes correspond to higher harmonics. This restricts our consideration by the condition $L_G \ll H$ (assumed  above).  
\begin{figure}[t]
\centering
\includegraphics[width=9.0cm]{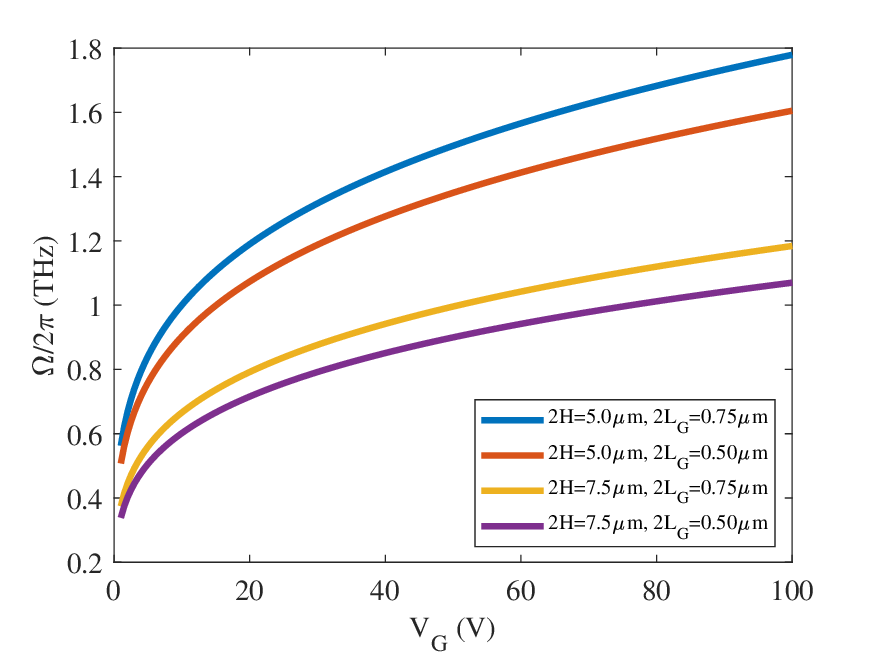}
\caption{Plasmonic frequency $\Omega/2\pi$ versus  bias voltage $V_G$ for GNR structures  width different length $2H$ , and widths $2L_G$ ($a = L_G/L = 2$).}
\label{Fig2}
\end{figure}

\begin{figure*}[t]
\centering
\includegraphics[width=15.0cm]{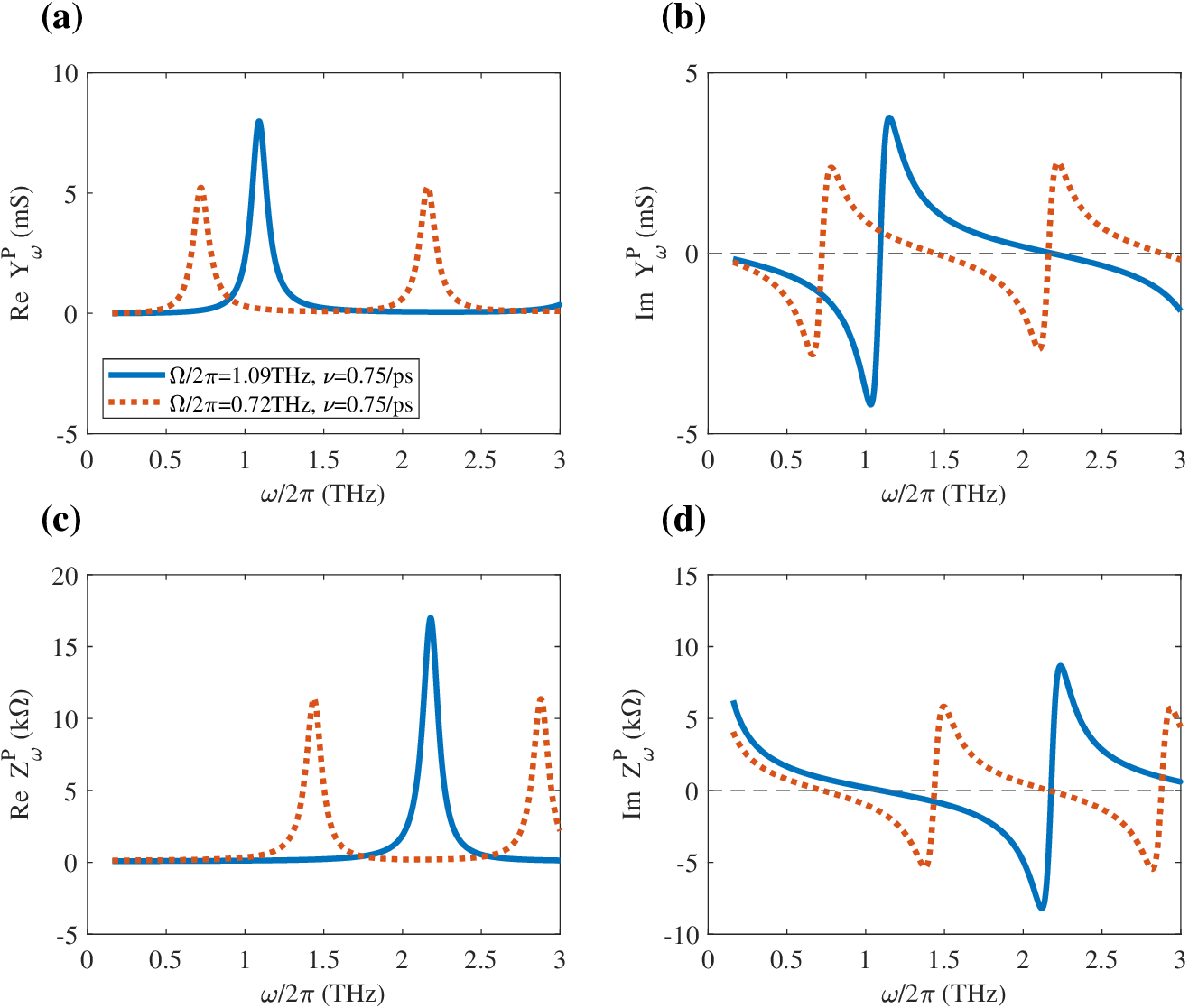}
\vspace{+ 5mm}
\caption{Frequency dependences of (a, b))  the real part  Re~$Y_{\omega}^P$, 
and  the imaginary part Im~$Z_{\omega}^P$
of the P-GNR admittance  and
(c, d) the real part  Re~$Z_{\omega}^P$, 
and  the imaginary part Im~$Z_{\omega}^P$
of the P-GNR impedance
   for different plasmonic frequencies $\Omega/2\pi$  ( $\nu = 1.0 $~ps$^{-1}$,  $a = L_G/L = 2$).}
\label{Fig3}
\end{figure*}

\begin{figure*}[t]
\centering
\includegraphics[width=15.0cm]{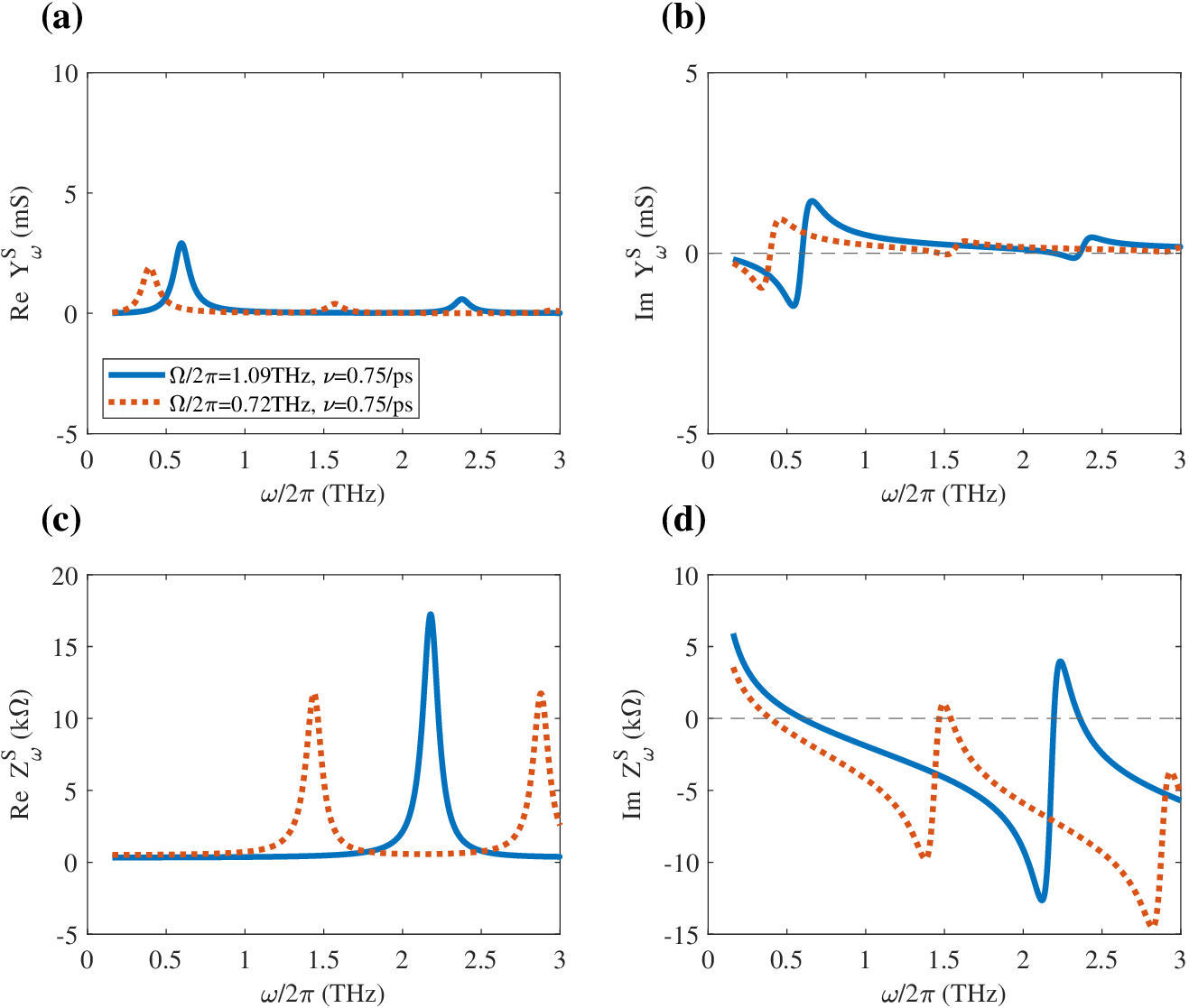}
\vspace{=5mm}
\caption{ The same as in Fig.~3, but for 
 the S-GNRs.}
\label{Fig4}
\end{figure*}

Figure~2  shows the dependences of the plasmonic frequency $\Omega/2\pi$ on the bias voltage $V_G$ for  different GNR geometries  calculated using Eq.~(13) with Eq.~(4). It is assumed that $\kappa_S = 4$, $H = 10~\mu$m, and $L_G/L = 2$. 
In line with Eq.~(12), an increase in the bias voltage $V_{G}$ leads to an increase in the characteristic plasmonic frequency $\Omega$ due to  the carrier density $\Sigma_G$ and the carrier Fermi energy $\mu$ rise when $V_G$ increases.  
The plasmonic frequency $\Omega$ is sensitive to the geometrical parameters $L_G$ and $L$, which determine the carrier density
and the GNR ac conductivity. 
The specifics of the GNR structures under consideration lead to relatively small inter-GNR capacitance and elevated values of the plasmonic wave velocity $s$. 
Consequently, rather
high  values of plasmonic frequency $\Omega$ (in the THz range) can be realized 
at fairly large lengths of the GNRs.

\section{GNR structure  admittance and impedance}

The net inter-GNR displacement current is given by

\begin{eqnarray}\label{eq14}
\delta J_{\omega} = \int_{-H}^Hdz \delta j_{\omega},
\end{eqnarray}
with the current density given by Eq.~(9) or Eq.~(10).
For the admittance of GNR structure under consideration 
$Y_{\omega} = \delta J_{\omega}\delta V_{\omega}$, we obtain (for the P- and S-devices, respectively)

\begin{eqnarray}\label{eq15}
Y_{\omega}^P= -i\frac{2HC{\omega}}{\gamma_{\omega}\,\cot\gamma_{\omega}}, \qquad
Y_{\omega}^S= -i\frac{2HC\omega}{\gamma_{\omega}(\cot\gamma_{\omega}-\gamma_{\omega})}.
\end{eqnarray}

The impedances of the P-GNRs and S-GNRs are given by 

\begin{eqnarray}\label{eq16}
Z_{\omega}^P = \frac{1}{Y_{\omega}^P},  \qquad
Z_{\omega}^S = \frac{1}{Y_{\omega}^S}.
\end{eqnarray}
At low signal frequencies ($\omega \ll \nu, \Omega$) the admittances and impedances of both GNR structures
tend to zero and infinity, respectively,  provided the inter-GNR leakage current is insignificant.
In the range of elevated frequencies comparable with the characteristic plasmonic frequency $\Omega$, the admittances and impedances exhibit the oscillations associated with the plasmonic response
of  the carriers in the GNRs.

Figures 3 and 4 show the P-GNRs and S-GNRs spectral characteristics. 

As follows from the analysis of the frequency characteristics of the P-GNRs described by the first  Eq.~(15),
Re $Y_{\omega}^P$ exhibits maxima at $\omega = \omega_n^P \simeq (2n-1)\pi\Omega/2$, where $n = 1,2,3,...$ is the plasmonic resonance index, i.e., when $\cot \gamma_{\omega_n^P} \ll 1$. At these frequencies, the peak height is proportional to the plasmonic oscillations quality factor $Q = 4\Omega/\pi\nu \gg 1$ being weakly dependent on the
resonance index. The first maxima of the curves  in Fig.~3(a) correspond to $\omega_1/2\pi \simeq = 0.72$ and 1.09 THz (dashed and solid lines). The second maximum for $\Omega/2\pi = 0.72$~THz corresponds to $\omega_2^P/2\pi \simeq 3\Omega/2\pi = 2.16$~THz.
Comparison of Figs.~3(a) and 3(b) shows that the maxima of Re $Y_{\omega}^P$ at $\omega=\omega_n^P$ correspond to a sharp change in the sign of Im $Y_{\omega}^P$ (so that Im $Y_{\omega_n}^P = 0$).
In contrast, Re $Z_{\omega}^P$ versus $\omega/2\pi$ dependences, given by the first Eq.~(16), exhibit maxima at $\omega = 2\omega_n^P$ [see Fig.~3(c)], i.e., at $\omega/2\pi \simeq (2n-1)\Omega/\pi$ ($\tan \gamma_{2\omega_n^P} \ll 1$) with the peak height proportional to $Q$.
Simultaneously, Im $Z_{\omega}^P$ turns to zero when $\omega = 2\omega_n^P$. This is in line with the plots in Fig.~3(d).

The frequency dependences of the S-GNR admittance shown in Fig.~4 differ from those for the P-GNRs.
This is due to different conditions of the plasmonic resonances. Indeed, considering the second Eq.~(15), one can see that the maxima of  Re $Y_{\omega}^S$ correspond to resonant frequencies $\omega =\omega_n^S$, which satisfy the condition
$\cot \gamma_{\omega_n^S} = \gamma_{\omega_n^S}$. This condition yields:
$\omega_0^S \simeq 1.72\, \Omega/\pi$ ("soft" mode) and $\omega_n^S \simeq  2\Omega(n  + 1/n\,\pi^2)$ with $n= 1,2,...$. 
As seen from Figs.~4(a) and 4(b), the positions of the  maximum of Re $Y_{\omega}^S$ and the point, where Im $Y_{\omega}^S =0$ are close  to $\omega_0^S/2\pi \simeq 0.4 $ (for $\Omega/2\pi = 0.72$~THz)
and to $\omega_0^S/2\pi \simeq 0.4 $THz (for $\Omega/2\pi = 1.09$~THz). 
As can be derived using the second Eq.~(15),
the height of the Re $Y_{\omega}^S$ maxima is proportional to 
$(4\Omega/\pi\nu)/[2 + (\pi\omega_n^S/2\Omega)^2]= Q/(2 + \pi^2n^2)$,
demonstrating a substantial decrease with increasing plasmonic index $n$ and the fact that the peak height is generally lower than the pertinent peaks of the P-GNR admittance.
Such a dependence of Re $Y_{\omega}^S$ on $n$ is clearly seen in Fig.~4(a).
 In particular, 
the ratio $\biggl( {\mathrm Re}~Y_{\omega_1^S}^S\biggr/{\mathrm Re}~Y_{\omega_0^S}^S\biggr) \simeq 0.23$.

It is interesting that in both P- and S-GNRs the Re $Z_{\omega}^P$ and $Z_{\omega}^S$ exhibit
 maxima at approximately the same frequencies having approximately the same height (compare Figs.~3(c) and 4(c).

One notes that the resonant frequencies at which Im~$Y_{\omega}^P$,  Im~$Y_{\omega}^S$, Im~$Z_{\omega^P}$, and Im~$Z_{\omega}^S$ equal to zero correspond to very low values of
Re~$Z_{\omega}^P$ and Re~$Z_{\omega}^S$.

The sharpness of the resonant peaks substantially depends on the carrier collision frequency $\nu$.
Their height is proportional to the plasmonic oscillations quality factor $Q \propto \Omega/\nu \propto M$, where $M$ is the carrier mobility in the GNRs.
A decrease in the mobility and, hence, in  the quality factor, 
leads to the pertinent  lowering of the resonant peak heights. This implies that the realization
of the GNR structures exhibiting pronounced plasmonic resonances requires using sufficiently perfect GNRs
or lowering the operation  temperature down from  room temperature leading to a substantial increase in the carrier mobility.

\section{Discussion and comments}
The structures based on  the vertically stacked  separately contacted GLs and GNRs
with the hBN, WS$_2$, Al$_2$O$_3$, and other insulating layers~
were studied and used in different devices (optical and infrared modulators and detectors,  THz detectors, THz detectors and photomixers, and others~\cite{36,37,38,39,40,41,42,43,44,45}.
The resonant plasmonic effects in such structures can lead to a substantial enhancement of  device characteristics.
The resonant plasmonic response in the lateral co-planar structures with  separately contacted GNRs, considered above,
is akin to that in the vertically stacked multiple-GL structures~\cite{40}.
However,
as predicted above, the lateral coplanar GNR structures can exhibit a pronounced plasmonic response controlled by the bias voltage even when the GNRs are
 relatively long (about several micrometers). 
 This is because the plasmonic frequency in such structures is determined by the inter-GNR geometrical capacitance.
This capacitance can be much smaller than the capacitance of the vertically stacked GLs.
Indeed, the  geometrical capacitance (per unit of the GNR length) of the lateral  structure with two coplanar GNRs considered above is given by 
$C = (\kappa/2\pi^2)c$, whereas the geometrical capacitance of the vertically stacked GNRs of the same lateral sizes  separated by the spacing $W$ is equal $C_V = (\kappa_SL_G/2\pi\,W) $, hence 

\begin{eqnarray}\label{eq17}
\frac{C}{C_V} = \frac{2cW}{\pi\,L_G}\frac{\kappa}{\kappa_S} = \frac{2cW}{\pi\,L_G} \frac{(\kappa_S +1)}{2\kappa_S}.
\end{eqnarray}
Comparing the plasmonic frequencies of co-planar and vertically stacked GNRs, $\Omega$  and $\Omega_V$, and  assuming that GNR lengths are equal to $2H$ and $2H_V$, respectively, 
 we obtain

\begin{eqnarray}\label{eq18}
\frac{\Omega}{\Omega_V}\biggr|_{\mu} = \frac{H_V}{H}\biggl(\frac{\pi L_G}{2cW}\frac{2\kappa_S}{(\kappa_S+1)}\biggr)^{1/2}
\end{eqnarray}
if the carrier Fermi energy $\mu$ (carrier density)  in the GNR structures is the same,
and

\begin{eqnarray}\label{eq19}
\frac{\Omega}{\Omega_V}\biggr|_{V_G} = \frac{H_V}{H}\biggl(\frac{\pi L_G}{2cW}\frac{2\kappa_S}{(\kappa_S+1)}\biggr)^{1/4}
\end{eqnarray}
for equal bias voltages, $V_G$, applied between the GNRs. 
Considering that in the devices with the vertically stacked GNRs, the thickness of the inter-GNR dielectric layer $W$ is relatively small  and setting 
$2L_G = 0.75~\mu$m, $L= 0.375~\mu$m ($c\simeq 2.36$), $W = (0.01 - 0.03)~\mu$m,  and $\kappa_S = 4$, we find
$(\Omega/\Omega_V)|_{\mu}\simeq (4.1 - 7.1)H_V/H$. This implies that in the coplanar GNR structures with the same plasmonic frequency can be realized at much longer GNRs ($H \gg H_V$).

As has been shown previously, the plasmonic resonances in the structures with the arrays of several vertically stacked GNRs (GNR stacks)  can be more pronounced~\cite{17,18,27}. The results obtained above
can be applied for the lateral arrays of vertically stacked GNRs
provided  the proper renormalization of the characteristic plasmonic frequency $\Omega$ given by Eq.~(12). In the case when the GNR stack comprises $K$  
non-Bernal stacked GNRs, the electron and hole density in each GNR is equal to $\Sigma_{G,K} = C\kappa\,V_G/4\pi\,e\,K$. Hence, for the carrier Fermi energy $\mu_K$, which determines the GNR Drude
 conductivity, we obtain $\mu_K \propto \sqrt {\Sigma_{G,K}} \propto 1/\sqrt{K}$. Accounting for that the net GNR stack conductivity is proportional to $K$, for the renormalized plasmonic frequency
 $\Omega_K$
 we find $\Omega_K \propto \sqrt{K}$.

 In principle, the plasmonic waves along the lateral  depleted channel formed in the graphene  p-n- junction predicted previously~\cite{46} resemble  those considered above. 
However, there
is a substantial difference between the plasmonic waves associated with the oscillating charges localized
in the isolated GNRs,  considered
by us, and the plasmonic waves associated with the residual charges in the depleted area in 
 graphene lateral p-n junctions.

 The voltage-controlled coplanar GNR plasmonic resonators can be used in the THz detectors exploiting
 the signal rectification in p-i-n diodes and the THz radiation sources based on the resonant-tunneling diodes. In the first case, an enhanced real part of the GNR structures admittance at the plasmonic resonances promotes the  rectified current amplification, which leads to an elevated detector responsivity. In the second case,  low values of the
 real part of the GNR  resonant impedance, can support  the THz oscillation self-excitation and radiation emission. 
 
Using  periodic coplanar S-GNR structures with multiple GNR pairs   can simplify the problem  matching these structures with the external circuits (e.g., with a load resistance or a THz antenna). These structures might be also  used for the THz photomixers exploiting  the
plasmonic resonances, GNR optical transparency, and strong light absorption in underlying layers, as well as for   the acoustic and photoacoustic  transducers. The resonant response associated with the excitation of plasmonic modes along the GNRs might result in a substantial increase in the impinging THz radiation.
This can be used for the performance enhancement of the bolometric THz detectors exploiting the carrier heating, in particular, their nonuniform heating (see, for example,~\cite{47,48,49,50}). 
 
\section*{Conclusions}
We explored the plasmonic response of the lateral coplanar GNR structures to the impinging radiation. The pertinent resonant frequencies are determined by the plasmonic frequencies, which  can fall into the THz range in  structures with fairly long GNRs
(about several micrometers). Due to the sensitivity of the plasmonic frequency to the bias voltage, the resonant response can be effectively voltage-controlled.
The coplanar GNR structures and their version with multiple interdigital GNRs can serve as the resonant cavities for different passive and active THz devices, including THz detectors,  photomixers 
using the interband transitions in the structure substrate (photoconductive antenna),
oscillators with the diodes exhibiting negative dynamic conductivity (for example, resonant-tunneling diodes), and the  acoustic transducers.

\section*{Acknowledgments}

The work at RIEC, FRIIS, and UoA was supported by the Japan Society for Promotion of Science (KAKENHI  Nos. 21H04546, 20K20349),
Japan.  The work at RPI was supported by AFOSR (contract number FA9550-19-1-0355).

\section*{Appendix A} 
\setcounter{equation}{0}
\renewcommand{\theequation} {A\arabic{equation}}

The signal components,  $\delta \Sigma^{\pm}_{\omega}$, of the hole ("+") and electron ("-") surface densities in the GNRs obey the continuity equations ($\partial.../\partial t = -i\omega...$)

\begin{eqnarray}\label{eqA1}
2L_Ge\frac{\partial \delta \Sigma^{\pm}_{\omega}}{\partial t} + \frac{\partial \delta J^{\pm}_{\omega}}{\partial z} =0.
\end{eqnarray}
Here 
\begin{eqnarray}\label{A2}
\delta J^{\pm}_{\omega} = -2L_G\sigma_{G,\omega}\frac{\partial \delta \varphi^{\pm}_{\omega}}{\partial z}
\end{eqnarray}
are the signal (linearized) currents along the GNRs, $\sigma_{G,\omega} =\sigma_G[i\nu/(\omega+i\nu)]$ is the GNR ac conductivity, $\sigma_G$ is the dc Drude conductivity, $\delta \varphi^{\pm}_{\omega}$
are the  signal potential spatial distribution along the GNRs, $e = |e|$ is the electron charge,
and $\nu$ is the carrier collision frequency.

The GNR ac charges per unit of their length $2L_Ge\delta \Sigma^{\pm}_{\omega}$  and the ac potentials $\delta \varphi^{\pm}_{\omega}$ are related to each other as
 \begin{eqnarray}\label{A3}
2L_Ge\Sigma^{\pm}_{G,\omega} = \pm C (\delta \varphi^{+}_{\omega} -\delta \varphi^{-}_{\omega}),
\end{eqnarray}
where $C$ is the geometrical inter-GNR capacitance. The latter equation corresponds to the
inter-GNR displacement current (per unit of the GNR length) $j_{\omega} = C\partial (\delta \varphi^{+}_{\omega} -\delta \varphi^{-}_{\omega})/\partial t$.

As a result, combining Eqs.~(A1) - (A3) for the signal components $\propto e^{-i\omega t}$, we arrive at Eqs.~(1) -(2).




\section*{Data availability}
All data that support the findings of this study are available within the article.

\end{document}